\documentstyle[prl,aps]{revtex}
\begin{document}

\title
{Electron-phonon interaction in a two-subband
 quasi-2D system in a quantizing magnetic field}

\author{V. M. Apalkov and M. E. Portnoi
\footnote{Corresponding author: Tel: +44-1392-264154; 
 Fax: +44-1392-264111; E-mail address: m.e.portnoi@ex.ac.uk}}
\address{
School of Physics, University of Exeter, Stocker Road, Exeter EX4 4QL, UK} 

\maketitle

\begin{abstract}

 We predict a double-resonant feature in the magnetic field
 dependence of the phonon-mediated mobility of a two-subband 
 quasi-two-dimensional electron system. These resonances take place when
 two Landau levels corresponding to different size-quantized subbands 
 are close to each other but do not coincide. We also discuss the 
 effect of non-equilibrium phonons. Rabi-like oscillations of electron
 population and emission of the phonons at new frequencies are predicted. 
\end{abstract}

 The specific feature of the two-dimensional (2D) electron system 
 in a strong magnetic field is its effective zero dimension
 due to Landau quantization of the in-plane electron motion. 
 This results in a strong suppression of electron-acoustic 
 phonon scattering because it is impossible for dispersionless 
 particles to fulfill simultaneously the energy and momentum 
 conservations under emission or absorption of a single phonon and 
 only the processes of the second order in electron-phonon interaction
 are allowed\cite{R1}. 

 The different situation occurs when the cyclotron energy is close
 to the intersubband splitting, where the subband structure is formed 
 by the quantization of electron motion in the direction perpendicular 
 to a confinement plane. In this case the second Landau level of the first 
 subband is close in energy to the first Landau level of the second subband. 
 When the energy separation between these levels becomes of the order of 
 $\Delta \sim \hbar s /l$ the transitions of electrons between 
 the levels with emission or absorption of a single phonon become 
 allowed,  where $s$ is the speed of sound and $l$ is the magnetic length.
 Such processes should increase the  dissipation conductivity, $\sigma _{xx}$. 
 In what follows we calculate $\sigma _{xx}$ to the first order in 
 electron-phonon interaction and show that it has a double peak structure 
 as a function of level splitting or a magnetic field.
 
 The above effect can be observed only in the system with electron 
 filling factor greater than unity. 
 Here and in what follows we neglect electron spins and consider a 
 spinless system. 
 We restrict our analysis to the case of the relatively low occupancy 
 of the intersecting levels so that we can consider electrons in 
 these levels as the non-interacting particles. 
 The high-occupancy case, in which the many-body effects become essential
 will be considered somewhere else\cite{apprb}.
 The many-electron effects in magneto-optics of two-subband system 
 were studied both theoretically and experimentally in Ref.\cite{hawr92}. 
   
 In a strong magnetic field the completely occupied first Landau level 
 of the first subband can be considered as a non-dynamical background. 
 Then the electron system can be mapped onto a two level system, 
 where  the first and the second levels are the second Landau level of
 the first subband and the first Landau level of the second 
 subband, respectively. The splitting $\Delta $ between the levels is 
 equal to 
$\Delta = \Delta _{01} - \hbar \omega _c$,
where $\Delta _{01} $ is the intersubband splitting and
$\hbar \omega _c $ is the cyclotron energy. For convenience we consider
 positive $\Delta $ only, the generalization to the case  of negative 
$\Delta $ is straightforward.

 We use the isotropic Debye approximation with a linear dependence of 
 the phonon frequency on its wave vector:
$\omega _{j} (Q) = s_{j} Q $,
where  $j$ is labeling the phonon mode,
 $j=1$ for longitudinal and $j=2,3$ for two transverse modes, 
$s_{j}$ is the speed of sound of the $j$th mode.

 The electron-phonon Hamiltonian can be written in the 
 form\cite{R2}:
\begin{equation}
H_{e-ph} = - \sum _{j, \vec{Q}} \frac{M_j (\vec{Q})}{\sqrt{V}} Z(q_z) \left[ 
   \hat{\rho }^{+} (\vec{q}) \hat{d}^{+}_j (\vec{Q}) + 
   \hat{\rho } (\vec{q}) \hat{d} _j (\vec{Q}) \right] \mbox{\hspace{3mm},}
\end{equation}
where the capital letters ($\vec{Q}$) mean the 3D vectors, their
projections onto the 2D layer are shown by corresponding small letters 
($\vec{q}$);  $\hat{d}^{+}_j$ is the creation operator of a phonon in the
$j$th mode, $V$ is a normalization volume, $\hat{\rho }^{+}(\vec{q})$ is the 
 electron density operator;
 $M_j (\vec{Q})$ are the matrix elements of electron-phonon
 interaction. 
 In GaAs these matrix elements have the form\cite{R2}:
\begin{equation}
\mbox{\hspace{-8mm}} M_j (\vec{Q}) = \sqrt{\frac {\hbar }{2 \rho_0 s Q}} 
  \left[ -\frac{eh_{14}}{\kappa \epsilon _0} 
           \frac{Q_x Q_y \xi _{j,z} +Q_y Q_z \xi _{j,x} +
                           Q_z Q_x \xi _{j,y}}{Q^2}
    - i \Xi _0 (\vec{\xi}_{j} \cdot \vec{Q} ) \right],
\end{equation}
 where $\rho _0 $ is the GaAs mass density, 
 $h_{14}$ and $\Xi _0$ are the parameters of piezoelectric
 and deformation potential couplings\cite{R3}, $\vec{\xi}_j$ is 
 the polarization vector of the $j$th phonon mode. 
 The form factor $Z(q_z)$ is determined by the electron
 spreading in $z$ direction:
\begin{equation}
Z(q_z) = \int dz e^{i q_z z} \chi_1(z) \chi_2(z) \mbox{\hspace{3mm},}
\end{equation}
where $\chi _1(z)$ and $\chi _2(z)$ are the wave functions associated 
with the first and second subbands, respectively. 
We use the Fang-Howard approximation\cite{R4} for these functions: 
\[
 \chi _1(z) = \sqrt{\frac{b^3}{2}} z \exp\left( -\frac{1}{2} bz \right) 
                                              \mbox{\hspace{3mm},\hspace{3mm}} 
 \chi _2(z) = \sqrt{\frac{b^5}{6}} z\left( z- \frac{3}{b} \right) 
                          \exp\left( -\frac{1}{2} bz \right)
                                              \mbox{\hspace{3mm}.}
\]
 The dissipative conductivity can be found from the equation:
\begin{equation}
  \sigma _{xx} = \frac{e^2}{2\pi l^2} \nu _{1} (1-\nu_2 )
   \frac{D_{12}}{kT} + \frac{e^2}{2\pi l^2} \nu _{2} (1-\nu_1 )
   \frac{D_{21}}{kT}
 = 2 \frac{e^2}{h}  \nu _{1} (1-\nu _2)
                      \frac{\hbar }{l^2} \frac{D_{12}}{kT} 
   \mbox{\hspace{1mm},}
\end{equation}
 where $\nu _1$ and $\nu_2 = \nu - \nu_1$ are the filling factors
 of the first and second levels, respectively; $\nu $ is the total 
 filling factor; $T$ is the temperature; $D _{ij}$ is the electron 
 diffusion coefficient corresponding to the phonon assisted transition 
 of the electron from level $i$ to level $j$. In expression (4) the 
 condition of thermal equilibrium $\nu _{1} (1-\nu_2 ) D_{12} = 
 \nu _{2} (1-\nu_1 ) D_{21} $ was used. 
 The diffusion coefficient is given by 
\begin{eqnarray}
D_{12} & = & \frac{l^4}{2} \left< \delta q^2 \right>  \nonumber \\
       & = & 
       \frac{l^4}{2} \frac{2 \pi}{\hbar }
      \sum _{j} \int \frac{d \vec{Q}}{(2\pi )^3} 
         \delta (\Delta  - \hbar s Q) n_{j}(\vec{Q})
            q^2  \left| M_j (\vec{Q}) Z(q_z) \right|^2 R_{01}(q)  
  \mbox{\hspace{3mm},}
\end{eqnarray}
where  $ n_{j}(Q)$ is the phonon distribution function, and
$R_{01}(q) = \frac{(lq)^2}{2} e^{-(lq)^2/2}$.
Equations (4) and (5) are derived using the arguments similar to those
in Ref.\cite{R5}.
 The filling factors $\nu_1$ and $\nu_2 $ are determined from the 
 standard Fermi distribution for the non-interacting two-level system:
\[
 \nu _1 = \frac{\nu }{ \nu +e^{-\Delta /2kT} 
    \left[ \xi +\sqrt{2-\nu +\xi ^2} \right] }
  \mbox{\hspace{3mm},\hspace{3mm}}
   \xi = (1-\nu ) \cosh(\Delta /2kT)  \mbox{\hspace{3mm},} 
\]
and $\nu_2 =  \nu - \nu_1 $. 

 Substituting equations (2), (3) and (5) into (4) and performing
 the integration we get the dissipative conductivity in the form:
\begin{eqnarray*}
\frac{\sigma _{xx}}{e^2 /h }
      & =  &   \nu_1(1-\nu_2) q_1^7 n(q_1) \int_0^1 du (1-u^2)^2 
                \left| Z(uq_1) e^{-q_1^2(1-u^2)/2} \right|^2 \times \\ 
      &    & \times  
           \left[ M_d + 
                 \frac{9}{8} \frac{M_{p,1}}{q_1^2} u^2 (1-u^2)
           \right]    
      +   \nu_1(1-\nu_2)   q_2^5 n(q_2) \frac{M_{p,2}}{8}  \times  \\
       &  & \times         \int_0^1 du (1-u^2)^3 (9 u^4 -2 u^2 +1) 
                 \left| Z(uq_2) \right|^2 
                  e^{-q_2^2(1-u^2)/2}    \mbox{\hspace{3mm},}
\end{eqnarray*}                      
where $q_1 =\Delta /\hbar s_1$, $q_2 =\Delta /\hbar s_2$
 and constants $M_d$, $M_{p,1}$ and $M_{p_2}$ are given by 
\[
M_d = \frac{1}{8 \pi}  \frac{\Xi ^2 }{kT \rho _0 s_1^2 l^3} 
\mbox{\hspace{2mm},\hspace{2mm}}
M_{p,1} = \frac{1}{8 \pi}  \frac{e^2 h_{14}^2 \hbar}{kT \rho _0 s_1 \
\kappa ^2 \epsilon _0^2 l} \mbox{\hspace{2mm},\hspace{2mm}} 
M_{p,2} = \frac{1}{8 \pi}  \frac{e^2 h_{14}^2 \hbar}{kT \rho _0 s_2 \
\kappa ^2 \epsilon _0^2 l}    \mbox{\hspace{2mm}.}
\]

 The typical double peak dependence of the dissipative conductivity 
 $\sigma _{xx}$ is shown in Fig.1a for $l(B_0)=50$\AA, $b=1/l(B_0)$ and 
 two values of the temperature: $T=3{\rm K}$ (dotted line) and $T=10{\rm K}$ 
(solid line), where we assume that the levels coincide at 
 magnetic field $B_0$. 
 The conductivity is shown for $\nu =0.1$ and it is normalized by 
 $\nu e^2/h$. 
 In these units $\sigma _{xx}$ has a weak dependence on the total 
 electron filling factor.  The two-peak structure is clearly seen 
 in the figure. The heights of the peaks increase with increasing
 the temperature which results from increasing the equilibrium 
 phonon density. 
 The described double peak structure should also be observed in 
 the phonon specroscopy\cite{R6,R7} experiments.

 We would also like to discuss another phonon-mediated effect, which 
 could be observed in the double-subband system under the influence
 of non-equilibrium ballistic phonons with the given direction of 
 momentum $\vec{e}_q=\vec{q}/q$. 
 This system is in fact a two-level system, in which the non-equilibrium
 phonons cause inter-level transitions between the states with 
 the momentum difference $q_0 \vec{e}_{q,||}$. Here $q_0 = \Delta/\hbar s$,  
 $\vec{e}_{q,||}$ is the projection of $\vec{e}_{q}$ on the 2D plane and we  
 use the Landau gauge with $y$-axis alone $\vec{e}_{q,||}$.
 Let us consider an electron, which is initially in the 
 state $\psi _{1, k_y}$ of the first level. 
 Absorption of a non-equilibrium phonon results in the transition 
 of this electron into the $\psi _{2,k_y+q_0e_{q,||}}$ state of the 
 second level.
 Then the stimulated emission of the phonon with the same $\vec{q}$ 
 transfers the electron back into $\psi _{1, k_y}$.  
 As a result the electron density oscillates between these two states with 
 the Rabi frequency\cite{R8}:
\[
\omega _R \approx n(q_0)/\tau (\vec{e}_q)
 \mbox{\hspace{3mm},}
\] 
 where $\tau (\vec{e}_q)$ is the time of spontaneous transition of an electron 
 from the state $\psi _{2,k_y+q_0e_{q,||}} $ into the state $\psi _{1, k_y}$.
 As in the usual optical Rabi effect one should expect the emission of the 
 phonons at the frequency $\Delta/\hbar$ as well as at two additional 
 frequencies:
 $\Delta/\hbar \pm \omega _R$. 
 In Fig.1b the frequency $\omega_0 =1/\tau (\vec{e}_q)$
 is shown as a function of level splitting. 
 The dependence has a double peak structure similar to the shape of the 
 dissipative conductivity.  

 To observe the phonon-induced Rabi splitting the initial distribution of 
 non-equilibrium phonons with the energy cutoff $\Delta _{cut}$ should be 
 created\cite{R8}, 
 where $\Delta < \Delta _{cut} < \Delta+\hbar \omega _R$ and 
 $n(q>\Delta _{cut}/\hbar s)=0$. 
 The interaction of such phonons with a 2D electron gas results in appearance
 of a pulse of the phonons with the energy  $\Delta + \hbar \omega _R$, 
 higher than in the initial phonon beam. 
 The length of this pulse is determined by the electron relaxation time, 
 which can be estimated from the dissipative conductivity: 
 $\tau _{rel}\sim e^2/kT\sigma _{xx}$. 
 The effect should be observable if the relaxation time is larger than the 
 period of oscillations of electron between two levels:
 $\tau _{rel} > 1/\omega _R$.
 
 In conclusion, we have shown that the dissipative conductivity of the 
 magnetically-quantized quasi-2D electron system with the filling factor 
 greater than unity has a double peak structure as a function of magnetic 
 field, when the cyclotron energy is close to the intersubband  splitting. 
 Such a system can be considered as an effective two-level 
 system and the peaks in conductivity result from the enhancement of 
 the  electron-phonon interaction when the separation between levels 
 becomes close to $\hbar s/l$. 
 A similar double peak structure should also be observed in the 
 non-equilibrium phonon spectroscopy, in which case phonon Rabi 
 sidebands are predicted. 

 This work was supported by the UK EPSRC.

\newpage

\begin{figure}
\begin{center}
\begin{picture}(110,70)
\put(0,0){\includegraphics{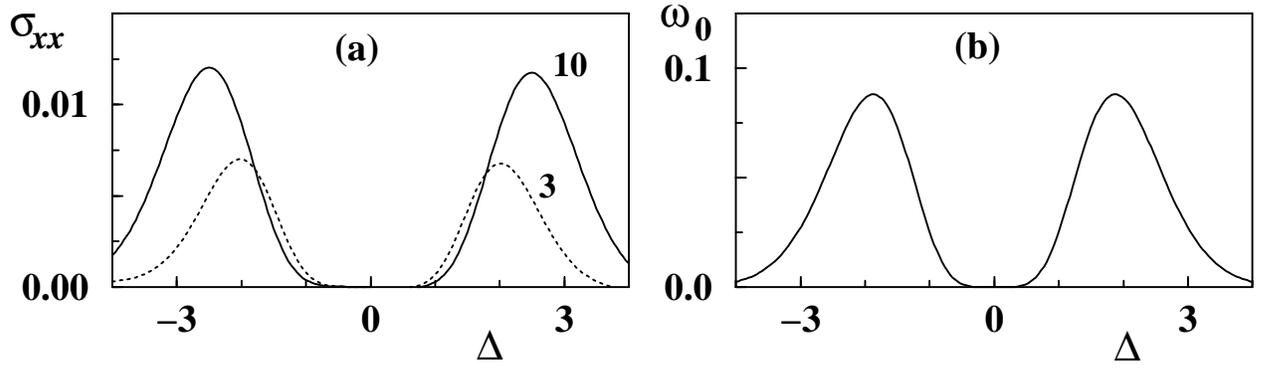}}
\end{picture}
\vspace*{7.0cm}
\caption{Dissipative conductivity $\sigma _{xx}$ (a) and frequency 
 $\omega_0 =\frac{1}{\tau (\vec{e}_q)}$ (b) as the functions of
 level splitting $\Delta $, where $\Delta $ is in  
 units of $\hbar s_1/l$, $\sigma _{xx}$ is in units of $e^2/h\nu$ and 
 $\omega _0 $ is in Kelvin. The data are shown for $l(B_0)=1/b=50$\AA. 
 The numbers near the curves in (a) show the temperature of the system 
 in Kelvin. 
}
\label{device}
\end{center}
\end{figure}

\end{document}